\documentclass{emulateapj}
\usepackage{graphicx}
\usepackage{amssymb}
\usepackage{epstopdf}
\usepackage{natbib}
\usepackage{color}
\usepackage{subfigure}
\usepackage{ulem}
\def\xmm{{\sl XMM-Newton}}
\def\chandra{{\sl Chandra}}

\def\suzaku{{\sl Suzaku}}

\def\kmps{\hbox{km $\rm{s^{-1}}$}}

\def\fexviii{Fe~{\sc xviii}}
\def\fexvii{Fe~{\sc xvii}}

\def\nvii{N~{\sc vii}}

\def\cvi{C~{\sc vi}}

\def\neix{Ne~{\sc ix}}
\def\nex{Ne~{\sc x}}

\def\oviii{O~{\sc viii}}
\def\ovii{O~{\sc vii}}

\def\mgxi{Mg~{\sc xi}}
\def\mgxii{Mg~{\sc xii}}

\def\sixiii{Si~{\sc xiii}}

\shorttitle{Charge-Exchange X-ray Emission from M82}
\shortauthors{Zhang et al.}

\begin{document}

\title{Spectral Modeling of the Charge-Exchange X-ray Emission from M82}


\author{Shuinai Zhang}
\affil{Purple Mountain Observatory, CAS, Nanjing, 210008, China}
\affil{Key Laboratory of Dark Matter and Space Astronomy, CAS, Nanjing 210008, China}

\email{snzhang@pmo.ac.cn}

\author{Q. Daniel Wang}
\affil{Astronomy Department, University of Massachusetts, Amherst, MA01003, USA}

\author{Li Ji}
\affil{Purple Mountain Observatory, CAS, Nanjing, 210008, China}
\affil{Key Laboratory of Dark Matter and Space Astronomy, CAS, Nanjing 210008, China}

\author{Randall K. Smith and Adam R. Foster}
\affil{Harvard-Smithsonian Center for Astrophysics, Cambridge, MA 02138, USA}
\and

\author{Xin Zhou}
\affil{Purple Mountain Observatory, CAS, Nanjing, 210008, China}
\affil{Key Laboratory of Radio Astronomy, CAS, Nanjing 210008, China}

\submitted{Submitted May 18, 2014. Accepted for publication in Astrophysical Journal, August 12, 2014}



\begin{abstract}
It has been proposed that the charge exchange (CX) process at the interface between hot and cool interstellar gases could contribute significantly to the observed soft X-ray emission in star forming galaxies.
We analyze the \xmm/RGS spectrum of M82, using a newly developed CX model combined with a single-temperature thermal plasma to characterize the volume-filling hot gas.
The CX process is largely responsible for not only the strongly enhanced forbidden lines of the K$\alpha$ triplets of various He-like ions, but also good fractions of the Ly$\alpha$ transitions of \cvi~($\sim 87\%$), \oviii~and \nvii~($\gtrsim $50\%) as well. 
In total about a quarter of the X-ray flux in the RGS 6-30 \AA\ band originates in the CX.
We infer an ion incident rate of $3\times10^{51}\,\rm{s^{-1}}$ undergoing CX
at the hot and cool gas interface, 
and an effective area of the interface as $\sim2\times10^{45}\,{\rm cm^2}$ that is one order of magnitude larger than the cross section of the global biconic outflow.
With the CX contribution accounted for, the best fit temperature of the hot gas is 0.6 keV, and the metal abundances are approximately solar.
We further show that the same CX/thermal plasma model also gives an excellent description of the EPIC-pn spectrum of the outflow Cap, projected at 11.6 kpc away from the galactic disk of M82. 
This analysis demonstrates that  the CX is potentially an important contributor to the X-ray emission from starburst galaxies and also an invaluable tool to probe the interface astrophysics.

\end{abstract}


\keywords{galaxies: individual (M82) --- galaxies: starburst --- X-rays : galaxies}

\section{Introduction}
Galaxy-scale outflows (sometimes called superwinds) driven by supernovae (SNe) are ubiquitous in starburst galaxies \citep[e.g.,][]{heckman90, heckman05}.
Such outflows represent a primary mechanism of stellar feedback, which
injects energetic and chemically enriched materials into the circum-galactic
medium (CGM) or even the intergalactic medium (IGM) \citep{veilleux05}.
The outflows consist of multiphases, including X-ray-emitting
thermal plasma \citep[e.g.,][]{bregman95,griffiths00}, warm ionized gas \citep{shopbell98}, neutral atomic/molecular gas, and dust \citep{lehnert99,taylor01}, as shown in M82 --- the target galaxy of the present study.

As a prototype starburst-driven outflow, M82 has received 
some of the most extensive studies of individual galaxies. The salient 
parameters of the galaxy are listed in Table~\ref{tab:info}.
The outflow from the nuclear region of 
the galaxy extends at least 3 kpc away on 
both sides from the galactic disk along its minor axis.
It is widely assumed that the outflow is driven by 
volume-filling hot plasma, while 
cold and warm gases are entrained within the flow \citep[e.g.,][]{chevalier85}.
Little is actually known about the origin of the emission lines
and hence about the properties of the hot plasma and its interplay with the 
cool components. The soft X-ray emission ($\sim$0.5-1 keV), in particular, 
may arise from either the volume-filling hot outflow \citep{fabbiano88, bregman95} or mostly from the cool-hot gas interface with a very low volume-filling factor \citep{strickland00a, strickland02}. Without separating these 
very different contributions, one cannot determine the energetics, as well as 
the thermal and chemical properties of the hot plasma, hence the potential 
role of the outflow in regulating the evolution of the galaxy and 
its environment.

High resolution X-ray spectroscopy offers an opportunity to greatly
advance our understanding of galactic outflows. Here we use the 
{\sl XMM-Newton} Reflection Grating Spectrometer (RGS)
observations of M82 to demonstrate this potential. 
Although the X-ray emission from the galaxy is extended, the large 
wavelength dispersion of the RGS still allows us to detect or resolve 
key individual X-ray emission lines or complexes, such as K$\alpha$ triplets
of He-line ions, with confusion far less than that in 
an X-ray CCD spectrum (e.g., extracted from \xmm/European Photon Imaging Camera (EPIC) data). 
Indeed, existing studies
with the RGS data have already demonstrated the diagnostic power of the 
line spectroscopy. In the analysis of a 50 ks RGS observation of M82, 
\citet{ranalli08} found that the \oviii~Ly$\alpha$ line and the \ovii~K$\alpha$ triplet cannot be appropriately accounted for even with a multi-temperature thermal plasma model and argued that the \ovii~triplet may indicate a significant
contribution from charge exchange (CX)~\citep{beiersdorfer03}.
This process occurs when an ion captures an electron from a neutral
atom or molecule. The captured electron tends to 
be in an excited state; the subsequent downward cascades can produce
emission lines with relative intensities distinctly different
from a collisionally excited plasma \citep{foster12}. \citet{liu12} 
quantified the CX contributions by analyzing individual line components
of the K$\alpha$ triplets of Helium-like O, Ne and Mg ions. 
Because of the electron capture, CX contributes to the line emission from lower ionization states
than in equilibrium thermal plasma;
in the case of multiple CX reactions, the emission appears at increasingly lower ionization states.
Therefore, the spectral analysis of the emission arising even partly from
the CX could lead to very misleading
results on the temperature and chemical properties of the plasma, 
if only its thermal (collisionally-excited) emission is considered.
CX may also contribute significantly to the 
enhanced X-ray emission from the so-called outflow Cap, projected at 11.6 kpc 
away from the galactic disk of M82 \citep{lallement04}. This X-ray enhancement coincides 
with an H$\alpha$-emitting filament \citep{devine99}, which is also seen  in 
ultraviolet emission due to the reflection by dust \citep{hoopes05}. Although
the X-ray CCD spectra obtained from \suzaku~and \xmm~observations are well characterized
by two-temperature components of optically-thin thermal plasma, it has been suggested that
some of the observed emission lines could be due to the CX~\citep{tsuru07}.

In this paper, we take a step further to model the entire RGS spectrum of M82, accounting
for both thermal and CX contributions simultaneously. We assume a volume-filling, 
single-temperature, optically-thin plasma in collisional ionization equilibrium, which is 
responsible for the thermal emission, 
and an interface of the plasma with cool gas, where the CX occurs.
This simple model is particularly motivated by the development of an 
integrated CX spectral code \citep{smith12}.
We check how well the RGS spectrum can be fitted with the model
and what constraints can be obtained on the effectiveness of the
CX process or the effective interface area,
as well as the properties of the thermal plasma. We further apply the model
to the analysis of an \xmm/EPIC-pn spectrum of the Cap region
to check the consistency of the model and to infer the dynamics of the outflow.

The organization of the paper is as follows: 
We describe the RGS data and model in sections 2 and 3, 
We present the fitting results
in section 4 and explore their implications in section 5, characterizing 
the key parameters of the hot gas 
outflow and its interplay with the cool gas.

\section{XMM-Newton data}

The present study uses the longest (104 ks) \xmm~observation 
(ID: 0206080101) of M82. After removing time intervals of high background 
with a counting rate above 0.2 counts s$^{-1}$ in the energy band 
above 10 keV, the remaining effective exposure is 50 ks.
We extract RGS spectra from the exposure, using the standard `rgsproc' script 
of SAS (version 12.0.0) and the 98\% cross-dispersion 
(roughly 2 arcmin) width of the point spread function. 
The zero point position of the dispersion is set at the center of 
the galaxy disk of M82 (Table~\ref{tab:info}). The script also produces a 
simulated background spectrum based on blank-sky observations.
We combine the RGS1 and RGS2 spectra into one spectrum by the `rgscombine' script, 
and regroup it with at least 20 photons per bin to 
allow for $\chi^2$ fitting, which uses the Interactive Spectral 
Interpretation System \citep[ISIS version 1.6.2,][]{houck02} that includes all Xspec models.

\begin{deluxetable}{lcc}
\tablecolumns{3}
\small
\tablewidth{0pt}
\tablecaption{Parameters of M82}
\tablehead{\colhead{Parameter} & \colhead{Value} & \colhead{Ref.\tablenotemark{a}}}
\startdata
RA(J2000.0) & 09:55:52.7    &  1 \\
Dec(J2000.0) & +69:40:46     & 1\\
Distance         &3.52 Mpc   & 2\\
Scale & 1.02 kpc/arcmin     & 2 \\
Redshift          & 0.000677  &  1 \\
Disk Inclination       & 81.5$^{\circ}$  &  3\\
Star Formation rate   &   $\sim5\,{\rm M_\odot\,yr^{-1}}$    & 4    \\
\enddata
\tablenotetext{a}{References: 1, NASA/IPAC Extragalactic Database (NED, http://ned.ipac.caltech.edu);
2, \citet{jacobs09}; 3, \citet{lynds63}; 4,  \citet{strickland09}.}
\label{tab:info}
\end{deluxetable}

\begin{figure*}[htbp] 
 \centering
       \includegraphics[angle=0,width=\textwidth]{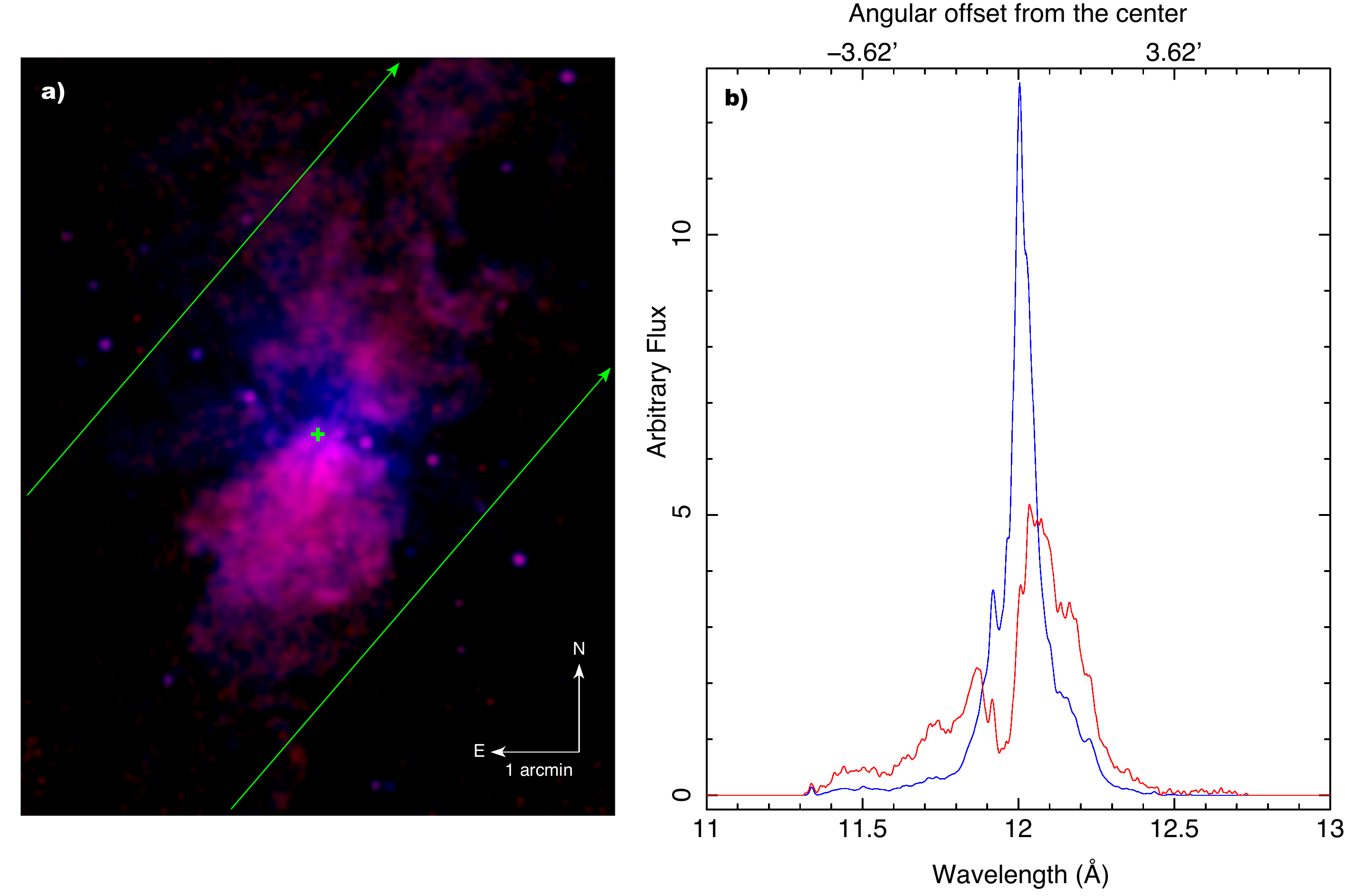} 
 \caption{(a) \chandra~images of the two bands, where the blue color marks the hard band (6-18 \AA) and the red color marks the soft band (20-30 \AA). 
The cross point is the center of M82. The RGS spectral extraction is between the two green lines, while the arrows mark 
the direction of the RGS dispersion. (b) Broadening profiles for a narrow Gaussian line (centered at 12 \AA~for an illustration), convolved with the two images. The blue line shows the profile from the hard band image, and the red line from the soft band image.}
 \label{fig:conv} 
\end{figure*}

The RGS spectrum covers the wavelength range of 6-30 \AA~(or 0.41-2.07 keV).
Since the RGSs are slitless spectrometers, the extracted spectra are a
convolution of the line spread function with the (projected) intrinsic spatial
distribution of the X-ray emission.
The dispersion of the RGSs is 0.138 \AA~per arcmin for a point-like source \citep{herder01}.
For the RGS observation with a position angle of 319.3$^{\circ}$, 
the dispersion direction is roughly along the long axis of the outflow, or 
the minor axis of M82, toward the northwest 
(Fig.~\ref{fig:conv}a).
As a result, the profile of an intrinsically narrow emission line (e.g., Fig.~\ref{fig:conv}b) is determined chiefly by 
the extension of the outflow. The blue and red sides of the line 
correspond to the northern and southern components of the outflow, respectively.
This line broadening is considered in the RGS spectral analysis (see \S~3.2).

We further extract an EPIC-pn spectrum of the Cap from the same \xmm\ observation. The data are processed with the standard 
procedures of SAS. Time intervals with counting rates above 0.7 counts~s$^{-1}$ 
in the energy band above 10 keV are removed to minimize the contamination of
high non-X-ray background, resulting in a total effective exposures of 51 ks.
For the spectral extraction, we adopt source and background regions that 
are nearly identical to those used by \citet{tsuru07} for the same 
observation. The extracted spectrum is grouped so that each bin contains a 
minimum signal-to-noise ratio of 3 in the 0.3-2 keV band.

\section{Spectral modeling}

Table~\ref{tab:pars} summarizes the model composition used to characterize the spectra of the 
outflow and the Cap. For the outflow, the model consists of: (1) a volume-filling, single-temperature 
plasma, which not only contributes to the optically-thin thermal emission, but also
sets the ionization states of ions before the CX occurs; (2) the CX emission
from the interface; (3) 
the continuum emission from point sources in the field, such as M82 X-1 \citep{matsumoto01}. 
The components (1) and (3) are 
represented by an {\small APEC} (Astrophysical Plasma Emission Code;
Foster et al. 2012) and a power law.
We fix the power law photon index as $\Gamma=1.6$ from fitting the EPIC-pn spectrum
of the central region of M82 \citep{ranalli08}, which is sensitive primarily to the 
energy range $\gtrsim 2$ keV (higher than that covered by the RGS data), 
where the confusion with the diffuse 
X-ray emission from the hot gas is minimal. 
The normalization of the power law is allowed to vary, depending on the 
specific field covered by the RGS spectrum. Key elemental abundances are also allowed to vary, relative to the solar 
abundances~\citep{anders89}.
In addition to a common, fixed Galactic absorption of $N_{\rm H}=4.0\times10^{20}\,\rm{cm^{-2}}$ \citep{dickey90},
intrinsic absorption is allowed for the (1)+(2) and (3) components separately.
This absorption is characterized by the model {\small TBABS} \citep{wilms00}, assuming the same solar 
abundances. For the Cap, no intrinsic absorption is assumed and the abundances
of the plasma are the same as those that best fits the outflow spectrum. 

In the following, we concentrate on
describing the newly developed CX spectral model and the approach used to account for the
line broadening due to the extended nature of the diffuse plasma.

\begin{deluxetable*}{ccc}
\tablecolumns{3}
\small
\tablewidth{0pt}
\tablecaption{Summary of spectral results}
\tablehead{\colhead{ } & \colhead{{\bf Outflow}} & \colhead{{\bf Cap}} \\
                  \colhead{Instruments } & \colhead{(RGS)} & \colhead{(EPIC-pn)} }
\startdata
XSPEC Model\tablenotemark{a}      & {\scriptsize A(G)[A(c) PL+A(o)(APEC\tablenotemark{*}+CX\tablenotemark{*})]} &      {\scriptsize A(G)(APEC+CX)}    \\
   \hline\noalign{\smallskip}
$N\rm{_H^G}\,\rm{(cm^{-2})}$  & \multicolumn{2}{c}{$4\times10^{20}$ (fixed)}    \\
$N\rm{_H^c}\,\rm{(cm^{-2})}$  &   $3.5(\pm0.3)\times10^{21}$   &  - \\
$N\rm{_H^o}\,\rm{(cm^{-2})}$  & $1.3(\pm0.2)\times10^{21}$  &  -  \\
PL photon index           &         1.6 (fixed)               &          -      \\
Norm (APEC)  &      $6.2(\pm1.2)\times10^{-3}$        &    $5.2(\pm0.8)\times10^{-5}$           \\
Norm (CX)       &      $2.1(\pm0.3)\times10^{-5}$        &    $4.1(\pm0.7)\times10^{-7}$         \\
$k_{\rm B}T$ (keV)                 &     0.60$\pm$0.01       &    0.58$\pm$0.09  \\
C 	& 0.41$\pm$0.37 	& 0.41 (fixed)\\
N 	& 0.85$\pm$0.17	&  0.85 (fixed)\\
O 	& 0.56$\pm$0.05	& 0.56 (fixed) \\
Ne    & 1.21$\pm$0.13      & 1.21 (fixed)  \\
Mg 	& 1.18$\pm$0.13	&1.18 (fixed)  \\
Si 	& 2.35$\pm$0.32      & 2.35 (fixed)  \\
Fe    & 0.31$\pm$0.03	& 0.31 (fixed) \\
\hline
 \multicolumn{2}{c}{Separate fitting abundance ratios\tablenotemark{b} } & \\
C/Fe  &  1.32$\pm$1.31     &  -  \\
N/Fe  & 2.74$^{+0.83}_{-0.59}$      &  -  \\
O/Fe  &  1.80$\pm$0.20    &  -  \\
Ne/Fe & 3.90$\pm$0.32     &  -  \\
Mg/Fe  &  3.81$\pm$0.53   &  -  \\
Si/Fe   &  7.58$^{+1.58}_{-0.84}$    &  -  \\
    \hline\noalign{\smallskip}
 $\chi^2$/d.o.f.        & 1.61 (1862.0/1158)   & 1.20 (69.8/58) \\
\enddata
\tablenotetext{*}{The models are convolved with line profiles generated from \chandra~images.}
\tablenotetext{a}{A(G), A(c), or A(o) stand for the absorption (TBABS) by the Galactic
foreground (G), gas intrinsic to the central region (c), or to the outflow region (o) of M82, with the 
corresponding column density of $N\rm _H^G$, $N\rm _H^c$, or $N\rm _H^o$.}
\tablenotetext{b}{From a separate fitting with the Fe abundance fixed to the above best fit value.}
\tablenotetext{note}{Chemical abundances are relative to solar values.}
\label{tab:pars}
\end{deluxetable*}

\subsection{CX modeling}

\citet{smith12} have presented an approximate model for the CX process and
the subsequent electron cascading down in energy levels, resulting in a 
predicted emission spectral model. While the details can be found in the work by
\citet{smith14} (or online\footnote{www.atomdb.org/CX/}), we here present 
a brief outline of the model.

The model assumes the most probable CX process that only a 
single electron is captured by an ion. The process can be represented by
$${\rm A^{q+} + N \rightarrow (A^*)^{(q-1)+} + N^{+}},$$
where a highly ionized ion ${\rm A^{q+}}$ (such as \oviii\ and \nex) picks up
an electron from a neutral species N (H, ${\rm H_2}$, and/or He), producing an excited ion ${\rm (A^*)^{(q-1)+}}$, 
which emits X-ray photons as it decays to the ground state. 

The emission spectrum depends on the relative $n$ and $l$ distribution of the
captured electrons, which affects the transitions during the subsequent
electron cascade to the ground state. 
The principal $n$ shell for the electron capture is determined from
Equation 2.6 of \citet{janev85}. The orbital angular momentum $l$ is
more complex: several distributions are available, each roughly
corresponding to different center-of-mass velocities of the system. At
lower collision energies, the separable $l$ distribution is suitable
\citep[Equation 3.59 of][]{janev85}.
Once captured, the electron cascades down to the ground state, purely 
by spontaneous or two photon emission. 
The model calculation includes 
a complete set of ions and radiative transitions based on the AtomDB 
database \citep{foster12}. Therefore, the calculated emission spectrum can 
then be directly compared to an observed one. 

We consider two extreme electron capture modes regarding to whether or not
an ion can have at most one CX. In one case, 
we assume that only one electron is captured.  
This is for CX in a sea of 
ions meeting occasional individual neutral atoms which have penetrated into 
the hot ``wind'', a circumstance suitable for the solar wind \citep{smith14}. 
In another case, the ``standard'' one, multiple CX captures occur rapidly until the ion becomes
neutral, an assumption justified by the large cross section for CX
- typically several orders of magnitude higher than the electron-ion
collisional excitation cross sections.
The difference between the two cases is mainly reflected by
the H- and He-like ion ratios (due to one or two CX captures) for C, N, O, and Ne.  
The prominent lines from Li-like ions are usually with energies lower than 0.3 keV, 
therefore are not in an RGS spectrum.

The cold gas in the outflow of M82 is either clouds entrained by the hot gas or the inflows due to the interaction with M81 \citep{melioli13}.
In the infrared band image of M82, the thickness of the filaments and shells of 
the cold gas are normally on the scale of several 10 pc \citep{engelbracht06}, much greater than the mean free path of the CX that occurs at the immediate vicinity of the interface due to its large
cross-section (typically a few $\times 10^{-15} {\rm~cm^2}$).
Thus, the situation in M82 is probably more suitable for the multiple CX case, in which the normalization parameter is physically meaningful as
$$\eta=10^{-5}f_{\rm H}/[4\pi D_{\rm A}^2(1+z)^2],$$
where $D_{\rm A}$ and $z$ are the angular diameter distance and the cosmic
redshift of the target, while $f_{\rm H}=\int{n_{\rm H}} v dS$
is the equivalent (metal abundance-dependent) hot proton incident rate (in units of $\rm{protons\,s^{-1}}$) through the interface. When the normalization 
is determined in a fit to the observed spectrum, $f_{\rm H}$ can then be measured.
Furthermore, if the proton density $n_{\rm H}$  of the hot plasma 
and the velocity $v$ (of ions relative to neutral atoms) can be estimated, one can then infer the
the effective area of the interface in the spectral extraction region. 

\subsection{RGS line profile modeling}
\label{ss:conv}

We use the Xspec convolution model {\small RGSXSRC} to account for 
the spectral broadening caused by the spatial distribution of the X-ray emission. 
We approximate the energy-dependent distribution using two high-resolution diffuse X-ray intensity images, 
constructed from mosaicing 30 \chandra~ACIS observations of M82 in the 6-18 \AA~(hard) 
and 20-30 \AA~(soft) bands and smoothed with a Gaussian filter of FWHM=5$'’$.
Figure~\ref{fig:conv}a shows that the X-ray intensity 
is quite concentrated (within the inner 1.5 arcmin
radius) in the hard band and is considerably more extended 
in the soft band. The lack of the emission near the galactic disk 
in this soft band is due to the strong photoelectric absorption
by the interstellar medium (ISM) in M82. 
The convolution of the {\small RGSXSRC} model uses 
the same dispersion center coordinates and position angle as for the RGS spectrum, as
well as an aperture radius of 4 arcmin, as suggested by 
\citet{ranalli08}, to cover the bulk of the emission from the outflow.

Figure~\ref{fig:conv}b demonstrates the spectral line broadenings as predicted from the above described convolution of a Gaussian emission line (12 \AA) with a turbulent velocity of 100 \kmps. The hard band image, where the thermal emission dominates, is used for the convolution with the {\small APEC} model.
The CX contribution\citep{liu12} becomes increasingly important and even dominates
at longer wavelengths. Thus the CX model is convolved with the soft band image. 
The dip near the center of this line profile reflects the absorption by the ISM.

\section{Results}

\begin{figure*}[htbp] 
 \centering
        \includegraphics[angle=0,width=\textwidth]{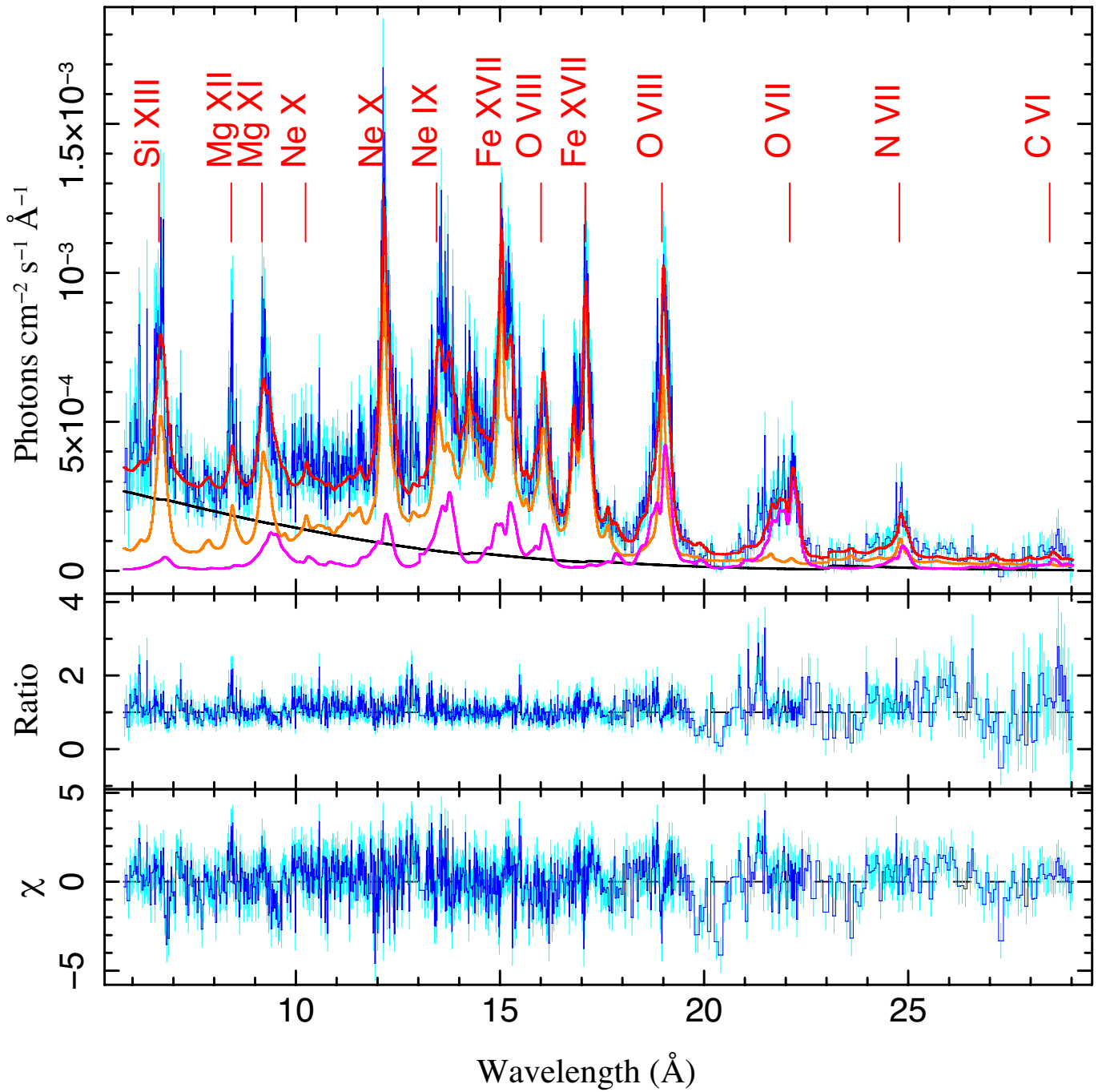} 
 \caption{RGS spectrum of M82 and the best fit model (red curve). 
The two lower panels show the data to model ratio and the data-model residuals (divided by
the counting error in each bin). 
The three emission components are power law (black), thermal plasma (orange), and CX (purple). }
 \label{fig:rgs} 
\end{figure*}

\begin{figure*}[htbp] 
\centering
      \includegraphics[angle=0,width=\textwidth]{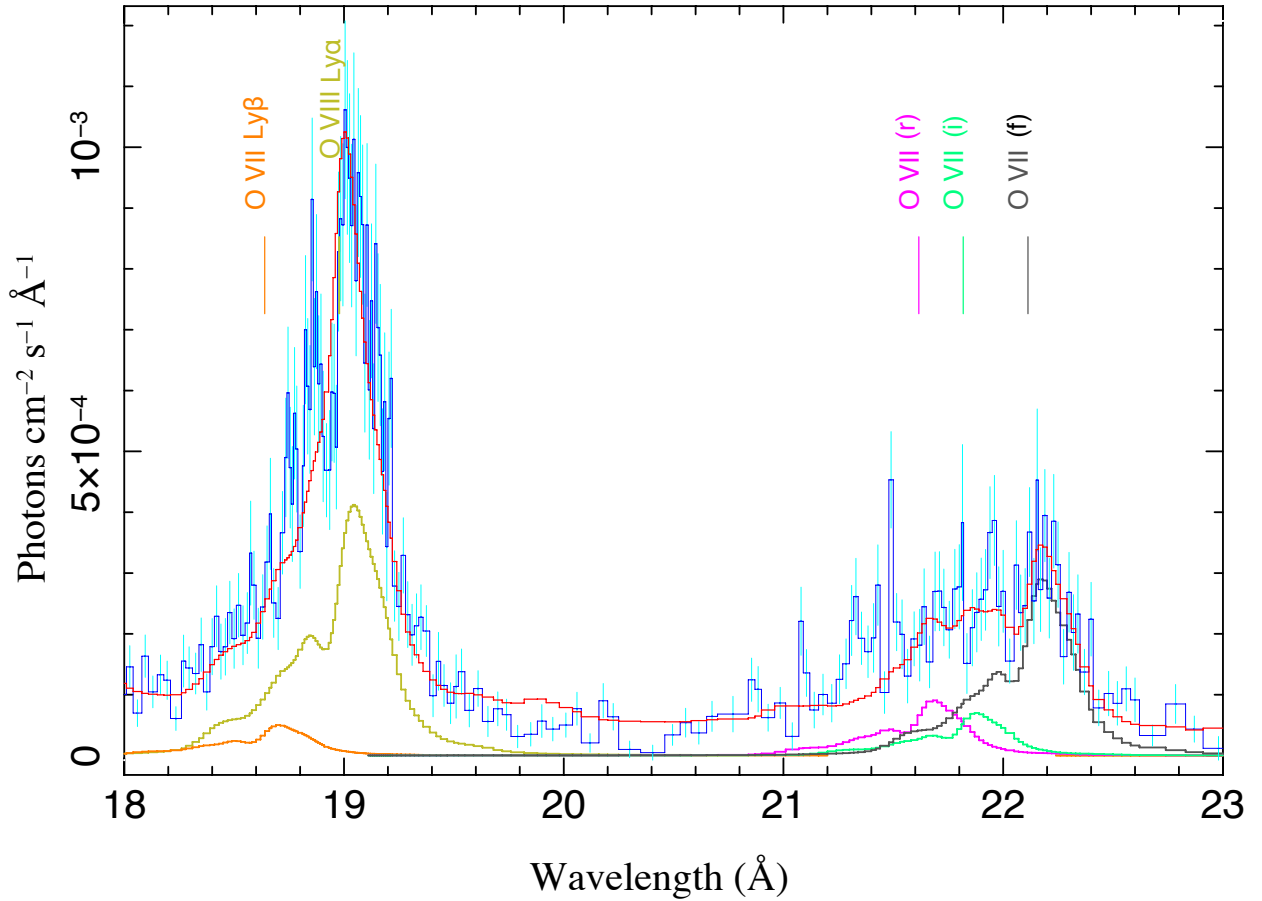} 
 \caption{Close-up of the RGS spectrum in the wavelength range covering  
\ovii\ and \oviii\ lines. The red curve is the total model prediction
as in Fig.~\ref{fig:rgs}. The CX contributions to various transitions 
are now plotted separately: the purple, green, and 
grey curves represent the resonance, inter-combination, and forbidden 
components of the \ovii\ K$\alpha$~triplet, respectively, whereas 
the orange and golden  curves  mark the contributions to 
the \ovii~Ly$\beta$ and \oviii~Ly$\alpha$ lines.}
\label{fig:cxtri} 
\end{figure*}

Figure~\ref{fig:rgs} presents the RGS spectrum of M82 and the best fit model.
A close-up of the \ovii\ and 
\oviii\ line complex is given in Figure~\ref{fig:cxtri}, showing the CX contributions to 
individual lines. In particular, the \ovii~forbidden line is well fitted by the CX 
contribution. A small residual at 21.4 \AA\ on the shorter wavelength
side of the \ovii(r) is probably due to uncertainties in the derived line profiles. 

Table~\ref{tab:pars} includes our best fit model parameters and their uncertainties 
at the 90\% statistical confidence level. 
The overall CX flux contribution is about one third 
of the thermal emission  in the 6-30 \AA~range and dominates at wavelengths $\gtrsim 20$~\AA. In Table~\ref{tab:flux}, we further compare the fluxes of the 
continuum, thermal, and CX components of the model. 
Systematic errors likely dominate in our simple model characterization of the X-ray spectrum.
To quantify such errors, we measured 
fluxes for major individual lines or complexes in the observed spectrum and compare the results with those in the best fit model (Table~\ref{tab:lines}). 
The CX to total line flux ratio ranges from 87\% for the \ovii~triplet to
less than 3\% for \fexvii~(16.9 \AA).
The `goodness' parameter represents the percentage of the line flux (in the observed spectrum) that is accounted for by the thermal+CX line component of the model.
The deviation from the ideal value (100\%) are all 
within 25\%, except for \mgxii~line (40\%). Considering the uncertainties in the
theoretical transition rates of the lines \citep[$\sim 30\%$;][]{guennou13} and the 
simplicity of our model, we conclude that the fitting is remarkably good
($\chi^2/n.d.f. = 1.61$, including only statistical errors).

\begin{deluxetable}{ccc}
\tablecolumns{3}
\small
\tablewidth{0pt}
\tablecaption{Fluxes of individual components}
\tablehead{\colhead{ Flux\tablenotemark{a}} & \colhead{{\bf Outflow}} & \colhead{{\bf Cap}} }
\startdata
 PL~(X-ray) & $7.3\times10^{-12}$  & - \\
 APEC~(X-ray) & $9.6\times10^{-12}$  & $8.2\times10^{-14}$\\
 CX~(X-ray) & $3.7\times10^{-12}$   &  $6.9\times10^{-14}$\\
 CX~(H$\alpha$)   &   $1.3\times10^{-12}$   &     $2.5\times10^{-14}$      \\
 Total~(H$\alpha$)    &   $4.6\times10^{-11}$ &    $1.5\times10^{-13}$    \\
    \hline\noalign{\smallskip}
Ion Incident Rate &    $3.2\times10^{51}$    &       $6.2\times10^{49}$     \\
\enddata
\tablenotetext{a}{The ion incident rate is in units of $\rm s^{-1}$, whereas the 
photon energy fluxes are all in units of $\rm erg\,cm^{-2}s^{-1}$. The X-ray 
fluxes are measured in the range of 6-30 \AA. The total H$\alpha$ fluxes from \citet{lehnert99} are included for comparison.}
\label{tab:flux}
\end{deluxetable}

\begin{deluxetable*}{lcc|ccc|r}
\tablecolumns{6}
\small
\tablewidth{0pt}
\tablecaption{Fluxes of key line features (in units of $10^{-13}\,\rm{erg\,s^{-1}\,cm^{-2}}$)}
\tablehead{\colhead{Line features\tablenotemark{a} ($\lambda$)} & \colhead{Range (\AA)}& \colhead{Data} & \colhead{Conti\tablenotemark{b}} &  \colhead{Thermal\tablenotemark{c}} & \colhead{CX}  & Goodness\tablenotemark{d}}
\startdata
\sixiii (6.65 \AA)	  & 6.4-7.0     & 9.56$\pm$0.38   & 4.52 & 5.24 (94\%)    & 0.31 (6\%)   &110$\pm$8 \%\\
\mgxii (8.42 \AA) & 8.3-8.6     & 3.27$\pm$0.14    & 1.45 & 1.00 (91\%)  & 0.10 (9\%)   & 60$\pm$5 \%\\
\mgxi (9.17 \AA)  & 8.9-9.6     & 7.17$\pm$0.19    & 2.82 & 3.00 (76\%)  &  1.19 (24\%)   & 96$\pm$4 \%\\
\nex 	(12.13 \AA) & 11.9-12.6  & 7.61$\pm$0.14   & 1.36 & 4.98 (81\%)   &  1.16 (19\%)   & 98$\pm$2 \%\\
\neix (13.5 \AA)  & 13.2-14.0  & 7.51$\pm$0.13  & 1.09 & 3.92 (66\%)    & 2.00 (34\%)   & 92$\pm$2 \%\\
\fexvii (15.1 \AA) & 14.7-15.5   & 7.33$\pm$0.13  &  0.84 & 4.78 (82\%)  & 1.05 (18\%)  & 90$\pm$2 \%\\
\oviii	(16.0 \AA)   & 15.7-16.4   & 3.26$\pm$0.09  & 0.57  & 2.23 (67\%) & 1.08 (33\%)    & 123$\pm$4 \%\\
\fexvii (16.9 \AA)  & 16.5-17.5  & 5.36$\pm$0.11   & 0.67 &  4.19 (97\%) & 0.12 (3\%)    & 92$\pm$2 \%\\
\oviii	(19.0 \AA)    & 18.5-19.5  & 4.54$\pm$0.10  & 0.44 & 1.71 (46\%)  & 2.01 (54\%)    & 91$\pm$2 \%\\
\ovii (22 \AA)        & 20.5-23.0  & 3.30$\pm$0.11   & 0.53 &  0.28 (13\%)&   1.84 (87\%)  & 77$\pm$3 \%\\
\nvii (24.8 \AA)     & 24.6-25.1  & 0.64$\pm$0.04   & 0.12  & 0.17 (42\%)&  0.23 (58\%)  & 77$\pm$6 \%\\
\cvi (28.5 \AA)      & 28.3-28.7  & 0.22$\pm$0.05   & 0.04  & 0.02 (13\%)&  0.13 (87\%)   & 83$\pm$23 \%\\
\enddata
\tablenotetext{a}{Identified with the dominant components; blending can be significant for some of the features: \nex(12.13 \AA) with \fexvii, \fexvii(15.1 \AA) with \oviii, \oviii(16.0 \AA) with \fexviii, and \oviii(19.0 \AA) with \ovii.}
\tablenotetext{b}{Power law flux plus the continuum flux of the thermal component.}
\tablenotetext{c}{Pure emission line flux of the thermal component and the fraction of its contribution to the emission line.}
\tablenotetext{d}{Percentage of the line flux in the observed spectrum explained by the model: (Thermal+CX)/(Data-Conti)}
\label{tab:lines}
\end{deluxetable*}

The metal abundances, except for Si and Fe, all seem to be consistent with being solar within 
a factor of 2, counting the statistical uncertainties (Table~\ref{tab:pars}).
The abundances of C, N and O are about 0.6 solar; Ne and Mg about 1.2, Si 2.4, 
and Fe is 0.3. These measurements do not seem to be sensitive to various potential systematic 
uncertainties. For example, fixing the power law index $\Gamma$ from 1.4 to 1.8
results in a change of individual elemental
abundance to be less than 0.1. Nevertheless, in the soft X-ray range considered here,
it is difficult to accurately determine the absolute abundances relative to H, because
H only contributes to the continuum. 
Instead, the data are more sensitive to the relative abundances among metal elements.
We thus also measure the abundances relative to Fe by setting their ratios as free parameters.
The resulting abundance ratios, included in Table~\ref{tab:pars},  are all supersolar, but within a factor of 4, 
except for the Si to Fe ratio, which is about 8 times higher.

\begin{figure*}[htbp] 
 \centering
       \includegraphics[angle=-90,width=5in]{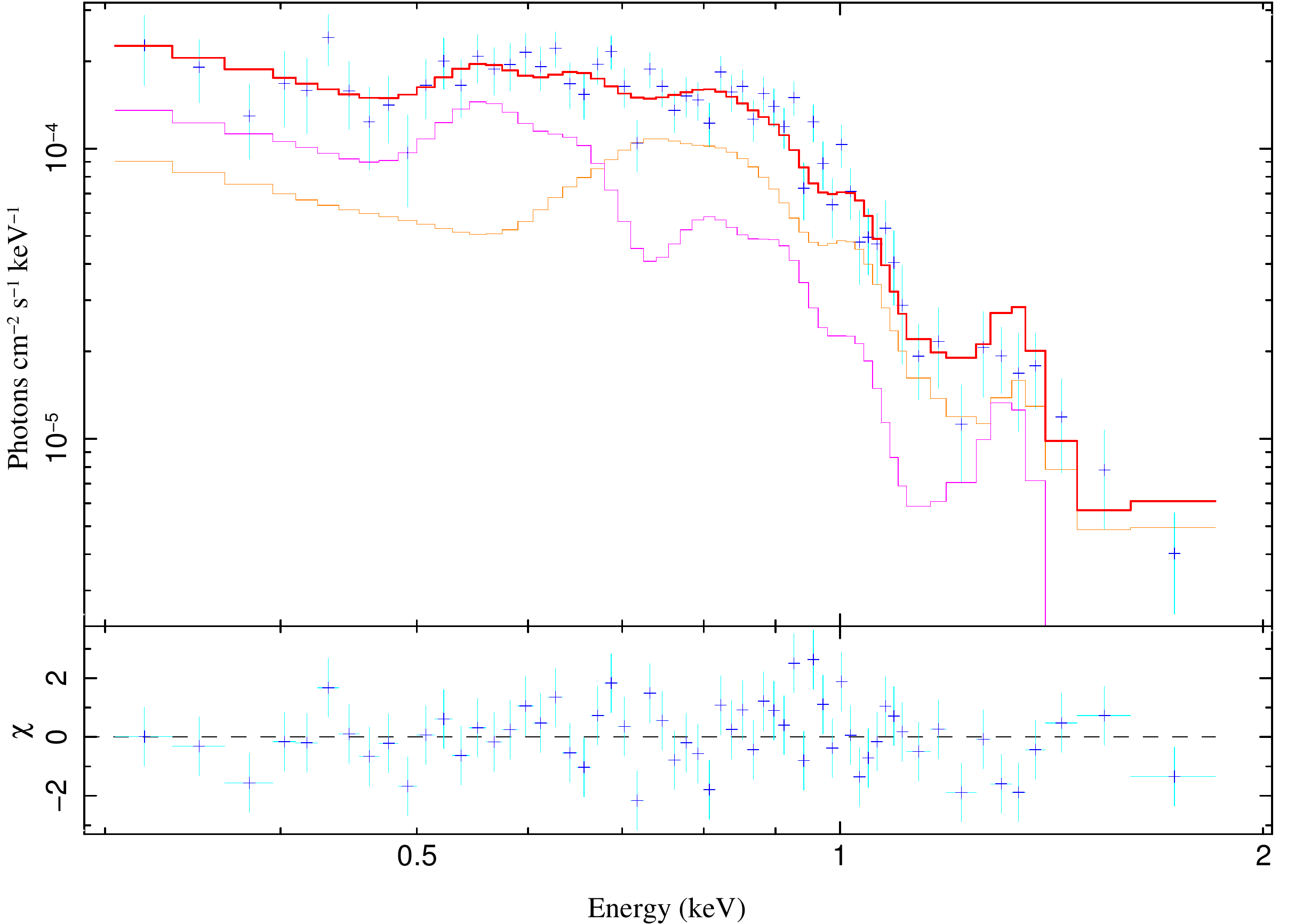} 
 \caption{EPIC-pn spectrum of the Cap and a fit with a model consisting of the APEC+CX components
renormalized from the best fit outflow model. The two components are
also shown separately: APEC (orange) and CX (purple).}
 \label{fig:cap} 
\end{figure*}

We find that the above APEC+CX model, which best fits the RGS spectrum of the outflow region,
also characterizes the pn spectrum of the Cap well. 
Because it is far away from the disk of M82, we exclude the power law and the intrinsic absorption.
The metal abundances are fixed to the best fit values of the 
RGS spectrum, which is reasonable because the X-ray emission at the Cap is presumably induced
by the outflow plasma from the vicinity of the galactic disk.
We first only allow the normalizations of the APEC and CX components to vary,  
which gives $\chi^2/d.o.f =1.20(70.6/59)$. 
We then also let the temperature of the plasma vary,
which yields slightly changes (well within the statistical uncertainties).
The results of this latter fit are included in Table~\ref{tab:pars}, and the plot is presented in Figure~\ref{fig:cap}.
The flux contributions from the APEC and CX components to the spectrum of the Cap 
are comparable to each other (Table~\ref{tab:flux}).

\section{Discussion}

The above results demonstrate an apparent success of the APEC+CX modeling of the 
line emission from the outflow and the consistency in explaining the overall X-ray spectrum of the Cap with the same model. In this section, we explore the implications of the results for 
understanding the hot plasma and its interface with the cool gas.
 
\subsection{Properties of the hot plasma}

The accounting for the CX contribution leads to an improved characterization of the thermal and chemical properties of the volume-filling hot plasma. 
So far the most sophisticated modeling of the RGS spectra without the CX uses thermal plasma with multiple or distributed temperature components.  
Close to the peak temperature of such modeling (\citealp[e.g., 0.7 keV from][]{read02} \citealp[or from][]{origlia04}; \citealp[0.5 keV from][]{ranalli08}) is our best fit effective temperature of 0.6 keV (with the CX accounted for). The CX process contributes more at longer wavelength, playing a role similar to the low temperature components in the plasma-alone modeling. 

Given the temperature, we can estimate the mass loading factor $\beta$ that characterizes the amount of mass loaded from ISM to the hot gas in units of the net mass injection rate of SNe and stellar winds (SWs) from massive stars.
The temperature of the mass-loaded hot gas is
$$T=0.4\frac{\mu m_{\rm H}\epsilon\dot{E}_{\rm SN+SW}}{k\beta\dot{M}_{\rm SN+SW}},$$
where $\mu m_{\rm H}\approx1.02\times10^{-24}\,\rm g\,cm^{-3}$ is the mean mass per particle (taking the solar elemental abundances), and $\epsilon$ is the mean thermalization efficiency of the kinetic energy released by SNe and SWs, assuming that the radiative cooling is not significant in the region \citep{mathews71}.
\citet{strickland09} calculated that the total SN plus SW energy and mass injection rates are $3.1\times10^{42}\,\rm erg\,s^{-1}$ and 1.4 $\rm M_\odot\,yr^{-1}$ respectively, assuming a continuous star formation with a  rate of 4.7 M$_\odot$ per year.
They also constrained the mean thermalization efficiency to be $30\%\leqslant\epsilon\leqslant100\%$ based on hydrodynamics simulations.
We adopt a middle value $\epsilon=70\%$.
As a result, a mass loading factor of $\beta\sim$10 is required to account for our inferred plasma temperature of 0.6 keV.
In comparison, the mass loading factor within the $\sim500$ pc starburst region is estimated as $1.0\leqslant\beta\leqslant2.8$ \citep{strickland09}.
Based on hydrodynamics simulations of the outflow, \citet{suchkov96} gives another estimate of the central mass loading factor as $\sim3-6$.  
If all these estimates are reasonable, then the comparison suggests 
that the mass-loading process occurs primarily in the path of the outflow within $\pm$3 kpc.
Alternatively, the low temperature plasma, as modeled here, may represent only part of the outflow, which has undergone substantial mass-loading (e.g., accompanied by CX).  The rest of the outflow may have had little mass-loading and may thus have too high (low) a temperature (density) to contribute significantly to the spectrum. In any case, our one-temperature APEC modeling is meant to provide only a simple characterization of the thermal plasma responsible for the observed soft X-ray emission via both collisional excitation and CX.

\begin{figure}[htbp] 
 \centering
       \includegraphics[angle=0,width=3.4in]{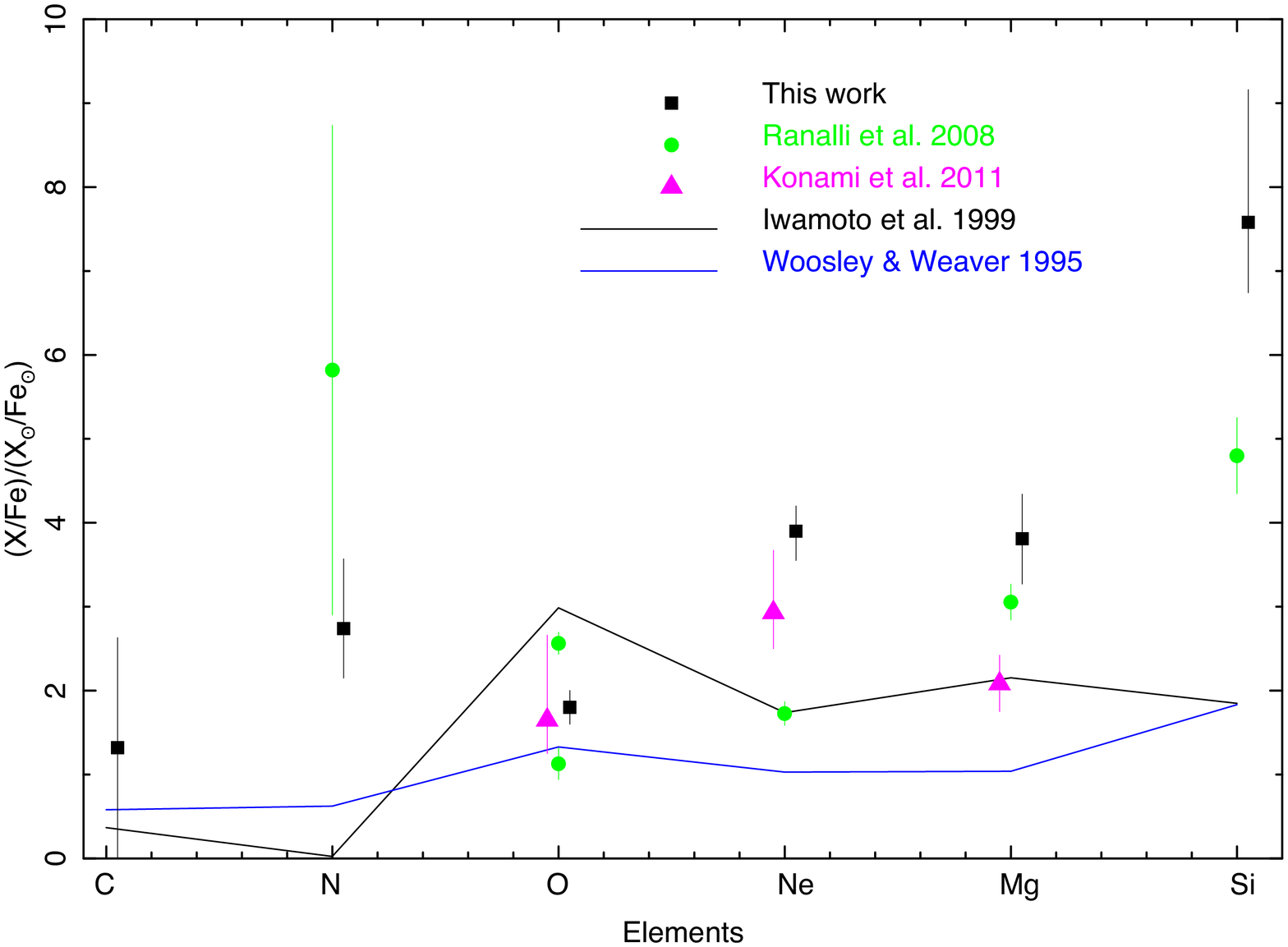} 
 \caption{X/Fe abundance pattern normalized according to solar values. The black squares mark measurements from this work, the green circles from the RGS study by \citet{ranalli08}, and the magenta triangles from the \suzaku\ study by \citet{konami11}. The black  line presents the predicted nucleosynthesis yields of type II SNe weighted with the Salpeter IMF in the range of 10-50 solar mass \citep{iwamoto99}, while the blue line shows the yields weighted at discrete stellar masses (11, 15, 20, 25, 30, 35 and 40 M$_{\odot}$; \citealp{woosley95}) but still according to the Salpeter IMF. }
 \label{fig:abundance} 
\end{figure}

Figure~\ref{fig:abundance} compares our measurements of the abundance pattern 
with those from RGS \citep{ranalli08} and \suzaku\ XIS \citep{konami11} studies of M82. 
Take the work by \citet{ranalli08} as an example. 
Their estimate of the N abundance or the upper limit of O is obtained from 
fitting the \nvii~(25 \AA) or \oviii~(19 \AA) line only, while the C abundance is not measured 
because of the poor counting statistics of the \cvi~(28.5 \AA) line.
Their lower limit of O is estimated from the plasma-alone modeling of the spectrum (6-18 \AA) including the \oviii~(16 \AA) line.
We have shown in the present work (Table~\ref{tab:lines}) that more than half of the line intensities
 of C, N, O (19 \AA) in the RGS spectrum can be attributed to the CX, leading to the reduced abundance estimates of these elements, compared to the results of \citet{ranalli08}.
The absolute Fe abundance that we have
measured (Table~\ref{tab:lines}) is lower than theirs again due to the removal of the CX contribution.
As a result, our Ne, Mg and Si to Fe ratios are higher than their results. 
Similarly, our ratios are all higher than those obtained from the analysis of the \suzaku\ XIS spectra~\citep{konami11} based on the thermal model consisting of three temperature components.  

For comparison, Figure~\ref{fig:abundance} also includes the abundance patterns expected 
from core-collapsed SNe in a starburst region.
The simulated type Ia SN nucleosynthesis yields \citep{nomoto97}
are not shown here, because as expected its X/Fe abundance ratios are much lower compared to our measurements (Fig.~\ref{fig:abundance}; Table~\ref{tab:pars}).
However, our measurements are also systematically higher than the predictions from type II SN yields weighted with the Salpeter initial mass function (IMF) in the range of 10-50 solar mass \citep{iwamoto99} or at discrete stellar masses (11, 15, 20, 25, 30, 35 and 40 M$_{\odot}$; \citealp{woosley95}).
The metal abundances prior to the last burst of star formation can be revealed by those measured for massive stars. 
The infrared measurements by \citet{origlia04} show that $\rm Fe/Fe_{\odot}=0.46^{+0.26}_{-0.17}$ and $\rm O/O_{\odot}=1.00^{+0.46}_{-0.32}$. 
Any enrichment by stellar winds and SNe would make the abundances higher. 
However, with the CX considered, we have
$\rm Fe/Fe_{\odot}=0.31\pm0.03$ and $\rm O/O_{\odot}=0.56\pm0.05$ (Table~\ref{tab:pars}).
Thus, our measurements indicate that
the O and Fe depletions may be important even in hot gas.
In particular, iron is believed to have the greatest fraction of its atoms depleted into dust grains in cold gas \citep{savage96}.
But dust grains are expected to be destroyed in shocks or sputtered gradually by collisions with electrons in hot gas. A substantial metal depletion in the hot plasma of M82 then suggests that a large fraction of its dust grains didn't pass a strong shock (e.g., only mass-loaded gently at the interface) and has not been in the hot phase for long (as may be expected for the outflow).

To estimate the density and energetics of the plasma, we need to know its volume.
The plasma is considered to be contained approximately in a bi-conic outflow
with a half cone opening angle of $\theta\sim30^{\circ}$ and a height of $\sim$3 kpc 
(Fig.~\ref{fig:conv}; \citealt{melioli13}), roughly corresponding to the enhanced X-ray region sampled by the RGS data. 
The total volume of the hot gas is then:
$$\int dV\approx\frac{2\Omega}{4\pi}\frac{4}{3}\pi R^3=(1-{\rm cos}\theta)\frac{4}{3}\pi R^3=5.6\times10^{65}\,\rm{cm^3},$$
where the $\Omega$ is the solid angle of each cone opening, and $R$ is the distance
from the galactic center to the top of each cone (assumed to be part of a ball).
The top surface area of the two cones is
$$S_{\rm bottoms}=\frac{2\Omega}{4\pi}4\pi R^2\simeq10^{44}\,\rm{cm^2}.$$

With the above volume estimate, we may now infer the
mean density of the plasma from 
the normalization of the APEC model: 
$$10^{-14}\int{n_{\rm e}n_{\rm H}} dV/(4\pi D^2)=0.0062,$$
where $n_{\rm e}\simeq 1.2n_{\rm H}$, and $D=3.52\,\rm{Mpc}$
(Table~\ref{tab:info}) is the distance  of M82.
Approximating the hot plasma to be homogeneous in the bi-conical outflow, 
we estimate $n_{\rm H}\simeq 0.04\,\rm{cm^{-3}}$. 
The total mass of the hot plasma is $M\simeq n_{\rm H}m_pV=2\times10^7\,{\rm M_\odot}$.
Considering the mass deposition rate of $\sim14\,\rm M_\odot\,yr^{-1}$ (\S~5.1),
the total mass suggests a period of about 1.4 Myr for the deposition of input materials.
Together with the temperature measurement (Table~\ref{tab:pars}), 
we infer the mean thermal pressure of the plasma
as $P/k\simeq2n_{\rm H}T=5.6\times10^{5}\,\rm{K\,cm^{-3}}$.
This pressure is about 2-20\% of the value
in the very central region ($<300$ pc) of M82, $P_{\rm center}/k\sim(0.3-3)\times10^{7}\,\rm{K\,cm^{-3}}$ \citep{bregman95},
and roughly corresponds to the pressure at the radius around 500 pc in the super-wind model \citep{chevalier85}.

\subsection{Hot and cool gas interface}

With the CX contribution characterized, we can estimate the total incident rate of ions onto the 
interface and its effective area in the outflow region. From 
the normalization of the CX model (Table~\ref{tab:pars}), we infer the
ion incident rate as $f_{\rm H}=\int{n_{\rm H}} v dS=3.2\times10^{51}\,\rm{s^{-1}}$,
or about 80 ${\rm M_\odot}$ hot ions per year.
Together with our model temperature, it implies that the CX involves a thermal energy of about $3k_{\rm B}Tf_{\rm H}\simeq10^{43} \rm~ergs~s^{-1}$ (including the contribution from electrons), which is comparable to the total expected mechanical energy input from SWs and SNe of about $10^{43} \rm~ergs~s^{-1}$ \citep{strickland09}.
This indicates that the CX is a manifestation of the substantial heat exchange between the hot and cool gases.

According to the conventional superwind modeling of the outflow from M82, 
the hot gas in the outflow region would have a velocity $\gtrsim 10^3$~\kmps, while the entrained cool gas 
moves slowly (e.g., $\sim$ 500 \kmps; \citealt{melioli13}). In reality,
the free expanding superwind modeling may work only partly in M82, perhaps in the northern side
of the galaxy \citep{liu12}, which is evidenced by the presence of the Cap \citep{lehnert99}. A good fraction 
of the hot plasma, especially in the southern side of the galaxy, 
seems to be morphologically enclosed by cool gas, 
as seen in images of H$\alpha$, dust radiation, and X-ray. Nevertheless, 
a good approximation of the relative velocity may be $v\sim$500 \kmps, comparable to the sound 
velocity of the hot plasma. Then with $n_{\rm H}\sim0.04\,\rm{cm^{-3}}$, as obtained above for the hot plasma, we estimate the total effective area of the interface as $\int dS \sim 2 \times10^{45}\,\rm{cm^2}$.
This area is about one order of magnitude larger than the summed cross-section of the two cones,
consistent with the turbulent nature of the interface around
cool gas clouds/filaments embedded in the hot plasma, and along the outer boundaries of the cones. 
The enlarged interface confines the outflow from free-expanding,
and makes the mass-loading into the hot gas efficient \citep{rogers13}.

How various potential astrophysical processes actually occur at the cool/hot interface remains somewhat uncertain. 
A layer of turbulent mixing may exist over the interface region driven by various instabilities, with a characteristic length scale of a few parsecs \citep{slavin93}. 
Also turbulence by its nature generates eddies cascading from macroscopic scales down to smaller ones. 
The processes such as CX and thermal conduction can operate efficiently on those microscopic scales.
For a typical CX length scale of $\sim10^{15}\,\rm cm$, the saturated thermal conduction by electrons may occur on time scales of about one year \citep{cowie77}. 
Given the relative velocity as 500 \kmps, the CX occurs on a similar time scale. 
Unlike thermal conduction, neutral atoms can move across magnetic fields 
or can thus be shaken out cool gas clouds in a highly turbulent environment, 
resulting microscopic mixing of gases in different phases.
In particular, since the ionization time scale of hydrogen is about ten years,
the CX can proceed efficiently.

\subsection{Cap as a discrete interface}

The Cap potentially provides the cleanest site to test the CX scenario and to probe the
properties of the outflow from M82. 
At a distance sufficiently far from the starburst
nucleus of the galaxy, this distinct feature should have a relatively uniform foreground 
absorption, which remains a systematic uncertainty in the analysis of the outflow region, and is largely 
free from contamination of discrete sources. Most importantly, the morphology of the reflected
UV radiation, as well as those of the X-ray and H$\alpha$ emissions, all indicate that
a discrete interface exists between the relatively isolated cool gas filament and the hot outflow, which is 
otherwise difficult to detect at the large distance from the galactic disk. 
The impact of the outflow on
the filament should naturally lead to the CX, as suggested by \citet{lallement04}. Although the existing
X-ray observations provide no distinct spectral line signature to quantify the CX contribution, 
the overall spectral consistency of the filament with  
the outflow is encouraging. The results from the Cap and outflow regions together can further be used to place 
additional constraints on the properties of the outflow at the distance of the Cap.

Similar to what we have done above for the outflow region near the disk (\S~5.2), 
we can infer the ion incident rate into the Cap as $6.2\times10^{49}\,\rm{s^{-1}}$ from the measured spectral normalization (Table~\ref{tab:pars}).
If we assume the outflow density of $\sim0.04\,\rm{cm^{-3}}$ and the absolute velocity of $\sim$1000 \kmps~at 3 kpc, plus the solid angle of the cone $\Omega=0.84\,\rm sr$, we may estimate the upper limit of the outflow ion particle flux as $\sim2.4\times10^{50}\rm~s^{-1}~sr^{-1}$. 
Together with the above ion incident rate and the off-galaxy center distance of the Cap as 11.6 kpc, we infer the lower limit of its interacting surface area as $\sim3\times10^7\,\rm pc^2$.
However, assuming the outflow-facing Cap to be approximately a circle of 1.8$^\prime$ radius 
 \citep{lehnert99}, the surface area would be roughly $10^7\,\rm pc^2$.
The discrepancy between the two areas indicates that either the Cap is elongated along our line of sight.

\subsection{Correlation between soft X-ray and H$\alpha$ emissions}

The presence of the CX naturally leads to the prediction of its contribution to the H$\alpha$ emission, in addition to the soft X-ray line emission that we have focused on. 
A hot proton undergoing a CX has a $\sim 20\%$ probability to produce an H$\alpha$ photon \citep{raymond12, lallement12}.
For the outflow region, the above estimated hot proton incident rate, $3.2\times10^{51}\,\rm{s^{-1}}$, can be translated into an H$\alpha$ flux of $0.43\,\rm{photon\,s^{-1}\,cm^{-2}}$, or $1.3\times10^{-12}\,\rm{erg\,s^{-1}\,cm^{-2}}$.
This CX contribution is about 3\% of the observed H$\alpha$ flux from the entire M82 \citep{lehnert99} and may be responsible for much
of the broadest H$\alpha$ line component with the measured FWHM = 333 \kmps~\citep[Table 2 in][]{westmoquette09}. 
This width is smaller than 
the thermal broadening of the 0.6 keV plasma with a FWHM of $\sqrt{8kT{\rm ln}2/m_p}=560\,\rm{km\,s^{-1}}$, where $m_p$ is the mass of proton.
But the velocity decomposition of the H$\alpha$ line can be quite uncertain 
(e.g., sensitive to the specific way in which the observed line is decomposed).
We may also expect that the shock driven into the cool gas from the impact of the high-speed outflow
be responsible for part of the H$\alpha$ emission,  
as well as the perhaps dominant contribution from the recombination of photoionized gas.
But still, the broadest H$\alpha$ component in the outflow region could be largely a result of the CX.
In the Cap region, the CX contribution increases to 17\% of the observed H$\alpha$ flux.

A spatial correlation between the diffuse X-ray and H$\alpha$ emissions was noted previously in M82 and other nearby star-forming galaxies \citep[e.g.,][]{strickland00b, strickland04}. This correlation
led to the impression that  both the soft X-ray emission and the H$\alpha$ emission may largely arise at boundaries 
between the hot and cold gases, occupying a small percentage of the space.
But the specific mechanism for such interaction is not clear, although shock \citep{strickland00a} and thermal conductions \citep[e.g.][]{dercole99} have been proposed.
Even with an outflow velocity of $\sim 10^{3}$~\kmps, 
it is not clear that the forward-shock propagating into the entrained dense cool gas would 
be energetic enough to produce X-ray-emitting plasma. 
Furthermore, it has been shown that the correlation is valid only globally; locally the X-ray enhancements can be anti-correlated with H$\alpha$ features \citep{liu12}.
This anti-correlation cannot easily be brought into a consistency with the shock heating interpretation.
 
In our model, about a quarter of the total observed X-ray emission in the 0.4-2~keV band arises from the CX emission at the interface. 
This specific mechanism provides a natural explanation for the global correlation between the soft X-ray  and H$\alpha$ emissions, 
while the local anti-correlation is expected for the X-ray emission from volume-filling thermal plasma enclosed by H$\alpha$-emitting cool gas.

\section{Summary and Conclusion}
We have made the first attempt to characterize the soft X-ray emission from M82 with a simple
physical spectral model. A novel component of this model is the implementation of the CX contribution, joined with a volume-filling, one-temperature, thermal plasma (APEC). This APEC+CX model is used
to fit both the entire RGS spectrum of the central outflow and the EPIC-pn spectrum of the Cap region. We summarize the main results and conclusions as follows:

\begin{itemize}
\item The model fits the RGS spectrum of the outflow well to an accuracy of 
matching individual emission features typically within 25\%. The fit shows that the CX is largely responsible for the enhanced forbidden lines of the K$\alpha$ triplets of various He-like ions, and contributes substantially to the Ly$\alpha$ transitions of \cvi~($\sim 87\%$), \oviii\ and \nvii~($\gtrsim $50\%).  About a quarter of the X-ray flux in the 6-30 \AA\ band originates in the CX. Furthermore, the required CX rate gives a direct estimate of the equivalent hot proton flux of $3.2\times10^{51}\,\rm{s^{-1}}$, which undergoes the CX at the hot and cool gas interface.  

\item The separation of the CX contribution improves the measurements of the thermal and chemical properties of the volume-filling plasma. The measured temperature of 0.6 keV is at
the high end of the range inferred from the previous modeling with thermal plasma only, albeit with multiple temperature components. The mean density of the plasma is $\sim$0.04 $\rm{cm^{-3}}$ and its total mass is $\sim2\times10^7\,{\rm M_\odot}$ within the inner $\sim 3$~kpc region. Our measured metal abundances are close to be solar (typically within a factor of $\sim2$) and show significant deviations from those based on thermal plasma-only modeling. The abundance pattern of X/Fe is systematically higher than the nucleosynthetic yields of core-collapsed SNe, suggesting Fe depletion onto dust grains even in the hot plasma.

\item A substantial mass-loading, a factor of $\sim 10$ of the SN and SW injected mass injection, from cool gas to hot plasma is inferred over the entire outflow region sampled by the RGS. This mass-loading is required to explain the low  temperature and high density of the plasma. The CX may just be a by-product of the mass-loading and may be responsible for much of the required heating of the cool gas, as indicated by a substantial energy flux involved in the process (comparable to the total mechanical energy input from the stars).

\item Assuming a characteristic relative velocity $\sim$500 \kmps~(comparable to the sound velocity of the hot gas) between the hot and cool gases, we estimate the effective area of the interface as $\sim2\times10^{45}\,\rm{cm^2}$, or one order of magnitude larger than the geometrical cross-section of the bi-conic outflows at  3 kpc off the galactic nucleus.
This large area of the interface is expected from the embedding of many cool gas clouds/filaments in the hot plasma and from the turbulent mixing between the two gas phases and can naturally accommodate substantial mass-loading to the hot plasma.

\item The EPIC spectrum of the Cap, as the most natural location for the CX process, can be well characterized by the same APEC+CX model (that best-fits the RGS spectrum of the outflow region) with a simple fitting to the normalizations. The CX contributes nearly half of the X-ray flux of the Cap in the 6 - 30 \AA~band, which requires an ion incident rate of $6.2\times10^{49}\,\rm{s^{-1}}$. The Cap may be elongated along our line of sight to account for this ion incident rate.

\item Our model naturally explains the global spatial correlation and some local anti-correlation between the X-ray and H$\alpha$ emissions.  The local anti-correlation,
apparent only in some regions, likely reflects the fact that the volume-filling hot plasma is surrounded and possibly enclosed by cool (H$\alpha$-emitting) gas across the bulk of the field. But the anti-correlation is weakened by the CX contribution to the X-ray emission, especially in the very soft X-ray band. On scales greater than individual bubbles/shells or other intermixing structures, the global correlation is expected. While the overall intensity of the H$\alpha$ emission is dominated by photoionization, plus shocking heating, the CX of hot protons with neutral atoms could still
be considerable (likely a few to twenty percent of the emission) and may be largely responsible for the broadest component of the observed H$\alpha$ emission line.

\end{itemize}

While the main goal of the present work is to demonstrate the potential utility of the CX spectral model in characterizing the soft X-ray line emission from a starburst galaxy, clearly more work needs to be done. 
In particular, we have only modeled the integrated RGS spectrum of the outflow. The unfavorable dispersion direction of the existing RGS observations does not allow us to probe the outflow structure along the galaxy's minor axis. Thus a deep \xmm/RGS observation with the dispersion perpendicular to the minor axis is highly desirable.
With imaging spectroscopic capability of upcoming X-ray detectors (e.g., {\sl Astro-H}), major advances can be made with the type of modeling illustrated here, which will lead to improved measurements of the plasma properties and unique 
insights into the interplay between hot and cool gases in such galaxies. 
Finally, we also plan to carefully examine scenarios other than the CX and the thermal plasma emissions (e.g., recombining plasma due to fast adiabatic expansion or to a recently extinguished active galactic nucleus), although they were not favored in brief explorations \citep[e.g.][]{liu12}.

\section*{Acknowledgements}
We gratefully acknowledge the anonymous referee for their comments that improved this paper. We thank John Houck for help on solving all ISIS software problems, and thank Keith Arnaud for help on the installation of the {\small RGSXSRC} script that is removed from the 12.7 version of Xspec. The work is partly supported by the National Natural Science Foundation of China under the grant 11203080, and by the Smithsonian Institution's Competitive Grants for Science. Li Ji is also supported by the 100 Talents program of Chinese Academy of Sciences.

\label{lastpage}

\end{document}